# Multi-scaling of Fractal Dimension and Relative Entropy as Diagnostic Tools for Irradiated Carbon Nanotubes

Shoaib Ahmad,[1,a,b,c] and Sumera Javeed [2,c]

[1]*National Center for Physics, QAU Campus, Islamabad 44000, Pakistan*

[2] *Pakistan Institute of Nuclear Science and Technology (PINSTECH), P. O. Nilore, Islamabad 45650, Pakistan*

Linear and nonlinear dissipative structures emerge in the irradiated single and multi-walled carbon nanotubes in the form of collision cascades and thermal spikes. These are diagnosed by the information-theoretic tools of fractal dimension and relative entropy by the probabilistic description of dissipative structures using the measure that depends upon the energy of the irradiating ion and the incremental energy step. Multi-scaling of the measure of the probability distributions induces variability of fractal dimension of the sputtered carbon atoms and clusters. Relative entropy of any two of the emitted components determines the nature and the extent of heterogeneity of the respective dissipative structures' probability distributions. By employing multi-scaling in going from relatively coarse-grain to finer scales, fractal dimension and relative entropy care shown to unambiguously distinguish and identify the information-generating collision cascades and thermal spikes that are created in the same spatial regions by the irradiating ion but in different temporal zones, in the mono- and multi-shelled carbon nanotubes.

**I. INTRODUCTION**

Experiments and simulations with irradiated carbon nanotubes reported during the past few decades have illustrated the importance and the utility of the radiation-induced physical processes, mechanisms and their effects at nanoscale. Irradiation of carbon nanotubes have been the subject of interest of a wide range of researchers from various disciplines as electron and ion beams can change and modify the structural and electronic properties[1-6]. Reviews and reports on single and multi-walled carbon nanotubes (SWCNTs, MWCNTs) irradiated with ion beams for inducing desired property changes have indicated vast potential for applications[7-9]. Computer simulations of energetic ions have studied creation of vacancy-interstitial pairs and the implanted ions in nanotubes[10-16]. Single vacancy and interstitial pair has been shown to be the prominent irradiation-induced mechanism. Multiple vacancies are predicted to be much less than single vacancies generated at all energies and with a variety of ions. However, a DFT-based calculation had indicated that the energy of formation of di-vacancy (DV) can be less than single-vacancy (SV) in SWCNTs and is a function of the nanotube diameter[13]. On the other hand, our Source of Negative Ions with Cesium Sputtering-SNICS[17]-based experimental results with $Cs^+$-irradiated nanotubes consistently demonstrated that multi-atomic clusters $C_x; x \geq 2$ have higher sputtering yields as compared with the monatomic $C_1$ yields, in a wide range of energies of irradiation[18-19]. We attributed the dominant cluster emissions to the nature of the dynamical processes that are initiated by an energetic $Cs^+$ ion and the topology of nanotubes is shown to favor the sharing of energy among the $sp^2$-bonded atoms of the adjacent hexagons-leading to the localized thermal spikes[20]. Nonlinear thermal spikes are demonstrated in this communication to be more efficient in sputtering clusters than the mechanism of monatomic sputtering by the collision cascades in the SWCNTs and MWCNTs.

---

[a] Author to whom correspondence should be addressed. Electronic mail: sahmad.ncp@gmail.com

[b] This research was performed while S. Ahmad was at CASP, Government College University, Lahore, Pakistan.

[c] S. Ahmad supervised the experiments and built the model and S. Javeed participated in experiments and data analysis.

The generation of collision cascades and thermal spikes is treated in this article, as the emergence of the local, dynamical events in the form of dissipative structures[21-23]. These dissipative structures occur in different time zones but share the same spatial region. The typical collision cascades time scales[8,10] are $\sim 10^{-14}$s while the localized thermal spikes last for longer durations that are equivalent to the atomic vibrational time scales $\sim 10^{-12}$s. In this communication, we analyze mass spectra of carbon atoms and clusters emitted as anions, from $Cs^+$-irradiated SWCNTs of 2 nm diameter and MWCNTs with $\sim 10$ nm diameter. The results are compared for the irradiations performed under similar experimental conditions for the two types of nanotubes.

Cs⁺ ion energy $E(Cs^+)$ with the incremental step $\delta E(Cs^+)$ is the basic measure of the dynamical system that comprises the initial input from the source of Cs⁺ into the sp²-bonded atoms of the hexagonal, curved, single- and multi-graphene sheets in the form of SWCNTs and MWCNTs. Information-theoretic descriptions of the emergent dissipative structures are explored by employing information-theoretic entropy-based diagnostic tools[24-28]. The dissipative structures emerge in the form of linear binary atomic collision cascades and nonlinear, multi-atomic thermal spikes that share the dissipated energy among atoms of the adjacent hexagons. When generated, both the mechanisms lead to the creation of vacancies that are composed of monatomic and multi-atomic vacant sites with the characteristic output signal in the form of the sputtered atoms and clusters. Multi-scaling of fractal dimension and relative entropy of the sputtered carbon species $C_x$ can unambiguously determine the nature of the underlying physical processes responsible for their emission. We show that there are significant differences in the generation of dissipative structures in the irradiated SWCNTs and MWCNTs. These can be explained by calculating the fractal dimension and evaluating the relative entropies of the emitted atomic and cluster species by coarse-graining through the choice of the relevant ionic energy steps $\delta E(Cs^+)$.

## II. Experimental method: mass spectra of $C_x^-$ sputtered from Cs⁺-irradiated nanotubes

Cs⁺-irradiation of single and multi-walled carbon nanotubes is performed in the Source of Negative Ions with Cesium Sputtering-SNICS. SWCNTs of 2 nm diameter and MWCNTs of nominal diameter of ~10 nm were compressed in Cu bullet targets for SNICS installed on 2 MV Pelletron at GCU, Lahore. The average length of the nanotubes ~8 − 10 microns. The sputtered atoms and the emitted clusters due to Cs⁺ irradiation acquire negative charge while leaving the target surface as a result of interactions with neutral Cs⁰ and are extracted as anions. A momentum analyzer was used to analyze mass spectra of the sputtered anions as a function of the cesium energy $E(Cs^+)$. The experiments were conducted by defining the Cs⁺ energy range $E(Cs^+)$ and choosing appropriate scale of $\delta E(Cs^+)$. By introducing multi-scaling of $\delta E(Cs^+)$ through appropriate selection of the measure $\zeta = \zeta(E(Cs^+), \delta E(Cs^+))$, the nature of the dynamical processes can be clarified. The atomic and cluster data from the irradiated SWCNTs and MWCNTs is used to evaluate the fractal dimensions and the relative entropies of the emitted species.

The normalized yields $N_{C_x}$ for C$_x$ with *x*-atoms as a function of $E(Cs^+)$ at a pre-determined scale of $\delta E(Cs^+)$ that can be varied, are obtained from the mass spectra by summing over the yields ($\Sigma N_{C_x}$) and then normalizing for each species; it provides $p(C_x) = N_{C_x}/\Sigma N_{C_x}$. This is the experimentally measured probability distribution of emission of the species C$_x$. Its variations as a function of Cs$^+$ energy and at the appropriately chosen scale $\delta E(Cs^+)$ defines the probability mass distribution $p(C_x)$ in eq. (1). We denote the probability as a function of the measure $\zeta = \zeta(E(Cs^+), \delta E(Cs^+))$ as

$$p(C_x) = p_{C_x}(E(Cs^+), \delta E(Cs^+)) \equiv p(\zeta) \tag{1}$$

The range of $E(Cs^+)$ is chosen between $E(Cs^+)_{min}$ and $E(Cs^+)_{max}$; where $E(Cs^+)_{min}$ is determined by the lowest detection limit of the experimentally measurable Cs$^+$-sputtered anion current. The lower limit $E(Cs^+)_{min}$ is 0.1 keV. $E(Cs^+)_{max}$ is determined by the maximum voltage of power supply (= 5 keV). We demonstrate in this communication that $\zeta = \zeta(E(Cs^+), \delta E(Cs^+))$ is the basic measure of multi-scaling in our experimental data. In the experiments reported here, $\delta E(Cs^+)$ is set at 0.1 and 0.5 keV. The lower limit is determined by the requirement of avoiding excessive Cs$^+$-induced damage in nanotubes[18,19]. The probability $p(\zeta)$ represents the combined effects of $E(Cs^+)$ and $\delta E(Cs^+)$ in eq. (1)

## III. Results

A carefully planned experiment generates the probability distribution as a function of the measure $\zeta = \zeta(E(Cs^+), \delta E(Cs^+))$ that depends on the ion energy $E(Cs^+)$ and $\delta E(Cs^+)$. The probability distribution $p(\zeta)$ is evaluated from experimental data for the sputtered species and used to derive a set of information-theoretic entropy-based parameters[27,28]. We make a distinction between fractals and multifractal before proceeding further. When we refer to the fractal dimension of a dissipative structure like collision cascade or thermal spike with reference to the emission of a particular species $C_x$ as $d_f^x$, it is implied that we are dealing with multifractals. Fractal has been defined as an object or a set while multifractal can be considered a measure[29-32]. Both, fractals and multifractal describe dissipative dynamical systems. When multi-scaling is employed, the probability distributions generate the information needed to specify, with the desired accuracy, the fractal or the Renyi information dimension[33-36]. Therefore, multifractal description of the set of the dissipative structures is based on multi-scaling, as opposed to the mono-scaling of the self-similar fractals[37]. We will refer to the multiple dissipative structures created by the irradiating ions in carbon nanotubes as multifractals where the dimensions of each structures are measured as a function of the measure $\zeta(E(Cs^+), \delta E(Cs^+))$.

A. **Coarse-grained measure** $\zeta = \zeta(E(Cs^+), \delta E(Cs^+) = 0.5\ keV)$

We present the experimental evidence of the atomic and cluster emissions from the two types of nanotubes with a coarse-grained measure of $\delta E(Cs^+) = 0.5\ keV$ and the range of $E(Cs^+)$ from 0.5 to 3.5 keV. With the comparatively coarse-grained measure of $\delta E(Cs^+)$, one can expand the $E(Cs^+)$ range to higher irradiation energies without excessive $Cs^+$-induced damage and avoid the effects of the Cs-implantation in nanotubes[18,19].

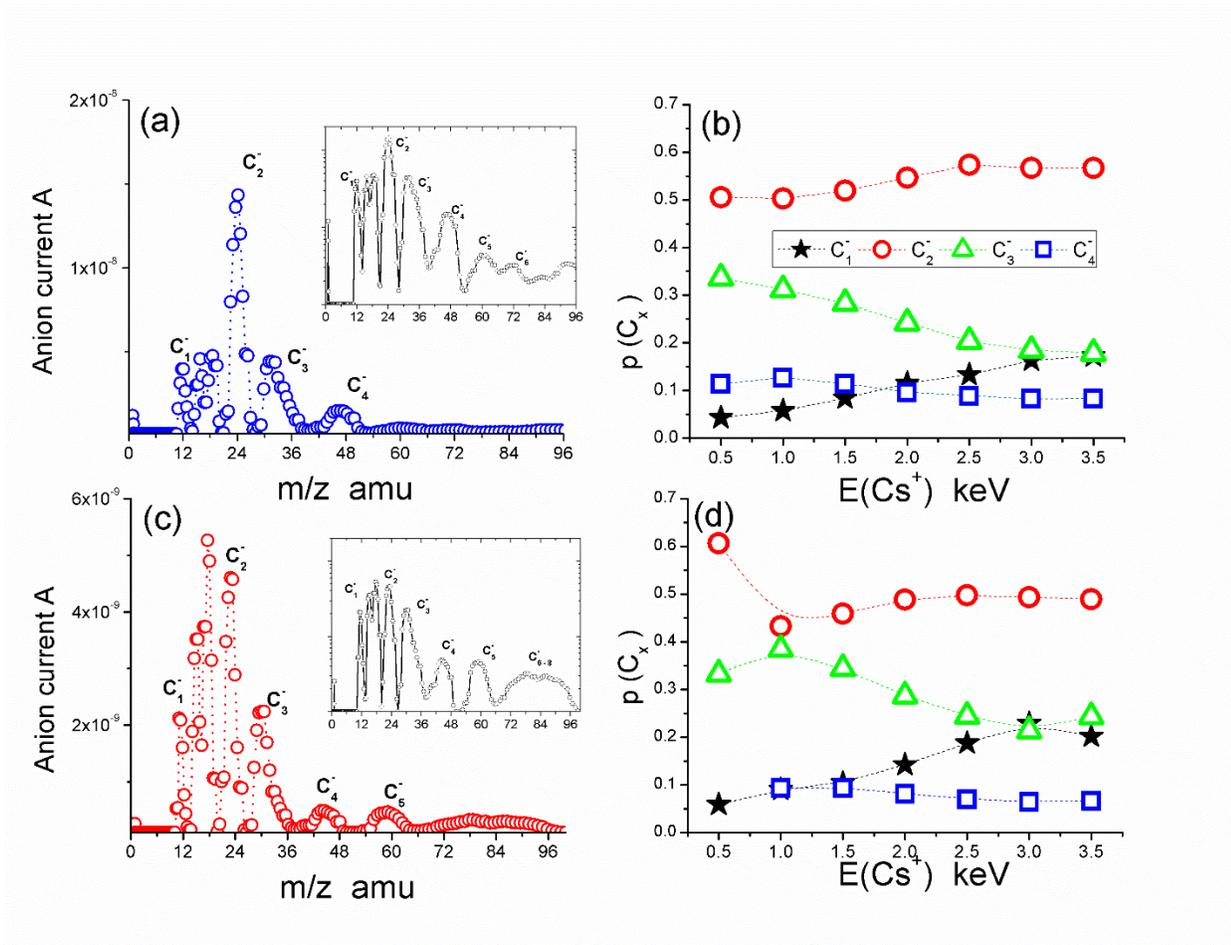

FIG. 1. $\delta E(Cs^+) = 0.5\ keV$. (a) Mass spectrum of anions sputtered from $Cs^+$-irradiated SWCNTs at $E(Cs^+)$ =1.5 keV. **Inset:** shows the log-plot of the anion currents to demonstrate the presence and dominance of large carbon clusters. (b) The probability of sputtering $p(C_x)$ of the four anionic species $C_1^-$, $C_2^-$, $C_3^-$ and $C_4^-$ are plotted as a function of $E(Cs^+)$. (c) Mass spectrum of sputtered anions from MWCNTs at energy $E(Cs^+)$ =1.5 keV. **Inset:** shows the log-plot of anion peaks. (d) The probability $p(C_x)$ for the four C species is shown for the same energy range as that of (b).

Figure 1(a) and (c) show two representative mass spectra of anions emitted from SWCNTs and MWCNTs at $Cs^+$ energy $E(Cs^+)$ =1.5 keV. In figure 1(b) and (d) the probabilities $p(C_x)$ of the four sputtered species $C_1$, $C_2$, $C_3$ and $C_4$ as anions, are obtained from their normalized emission densities, and plotted against $E(Cs^+)$. The figure clearly demonstrates the predominant emission of $C_2$, $C_3$ and $C_4$ while there is relatively lower probability of the emission of $C_1$. The cluster emission does not explicitly depend on ion energy while for $C_1$ the probability $p(C_1) \propto E(Cs^+)$; as one would expect from linear atomic collision cascade theories[29,30]. The non-dependence of $p(C_x); x \geq 2$ on $Cs^+$ energy variations can be traced in space-filling, multifractal, localized thermal spikes[32].

B. **Fine-grained measure** $\zeta = \zeta(E(Cs^+), \delta E(Cs^+) = 0.1\ keV)$

The fine-grained momentum analyzed spectra of $C_x^-$ anions sputtered from SWCNTs and MWCNTs are shown in Fig. (2) for the $E(Cs^+)$ range 0.3 to 1.0 keV with $\delta E(Cs^+)$ = 0.1 keV. The most significant anionic peak with the highest peak current is due to the diatomic cluster $C_2$. It is the omnipresent species at all energies and in the spectra from both types of carbon nanotubes as shown in Fig. 2(a) and (b). The other important feature is the ratio of the areas under peaks of $C_2^-$ and $C_3^-$ that remains approximately constant throughout the irradiations with energies $E(Cs^+)$ in the range from 0.3 to 1.0 keV for SWCNTs and MWCNTs. Monatomic carbon $C_1^-$ shows an energy dependent profile where $p(C_1) \propto E(Cs^+)$. This aspect of atomic sputtering has generally been discussed based on single vacancy generation[10,11,29,30]. The other noticeable feature is the presence of water-related peaks between $C_1^-$ and $C_2^-$. More pronounced water-related peaks are visible in the spectra from MWCNTs than those emitted from SWCNTs. This is due to the possibility of larger areas for water accumulation on MWCNTs as compared with SWCNTs. The sputtering of water from the irradiated nanotubes is discussed elsewhere[33].

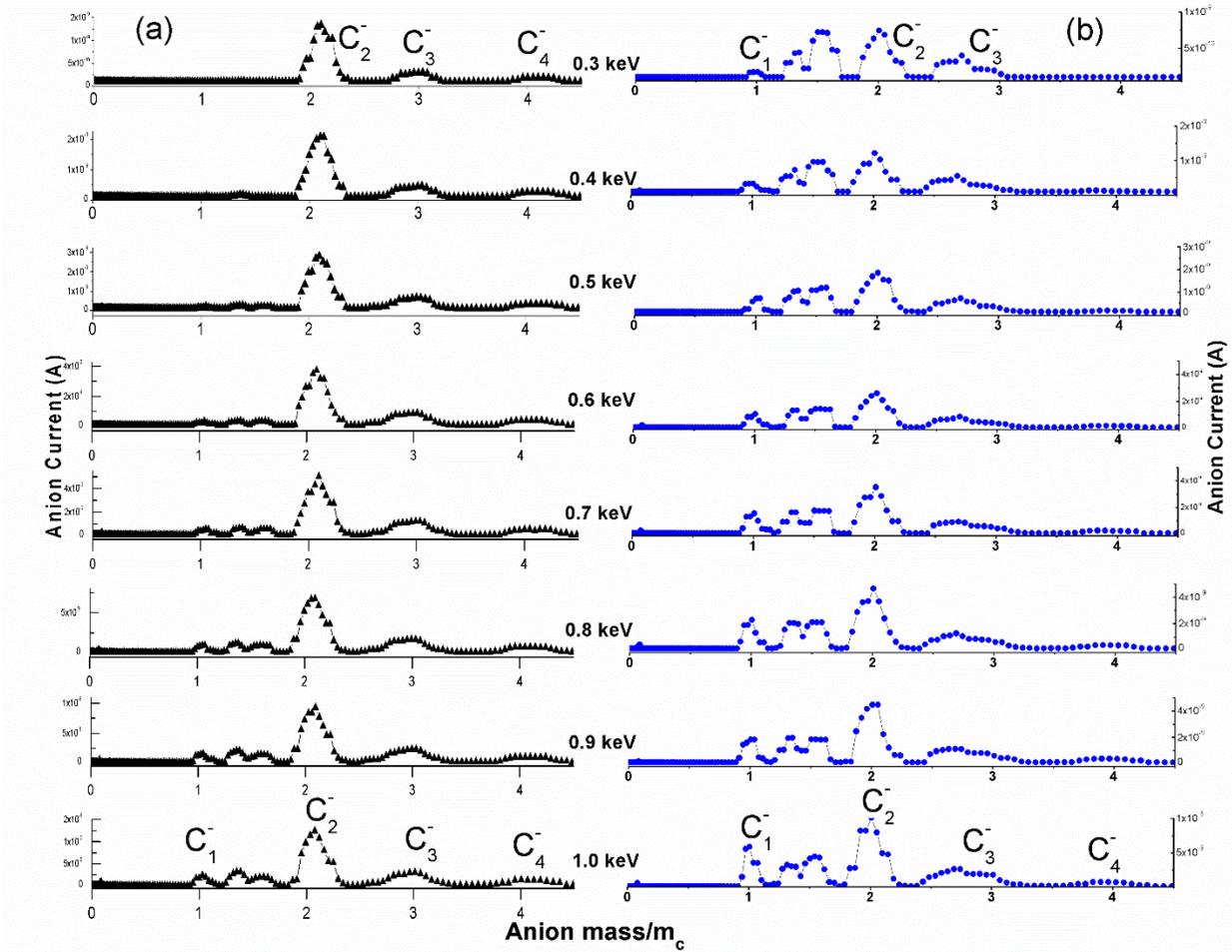

FIG. 2. $\delta E(Cs^+) = 0.1\ keV$. Anions $C_1^-$, $C_2^-$, $C_3^-$ and $C_4^-$ are the constituents of the mass spectra emitted from $Cs^+$- irradiated SWCNTs in Fig. 1(a) and MWCNTs in Fig. 1(b). Eight spectra from each nanotube are shown for $E(Cs^+)$ = 0.3 to 1.0 keV with $\delta E(Cs^+)$ = 0.1 keV. $C_2^-$ and $C_3^-$ are the major sputtered species with gradually increasing $C_1^-$ content with increasing $E(Cs^+)$. Water-related peaks are present between $C_1^-$ and $C_2^-$ with varying number densities.

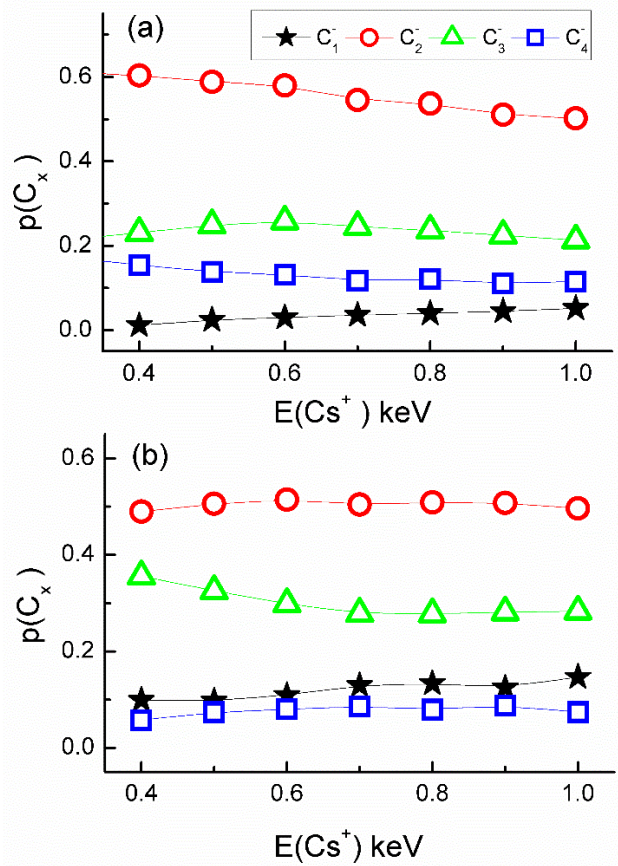

FIG. 3. $\delta E(Cs^+) = 0.1\ keV$. Normalized emission probabilities $p(C_x)$ for $E(Cs^+) = 0.3\ to\ 1.0\ keV$ at $\delta E(Cs^+) = 0.1\ keV$ of $C_1^-$, $C_2^-$, $C_3^-$ and $C_4^-$ are plotted as a function of $E(Cs^+)$. The data for SWCNTs is shown in 2(a) and for MWCNTs in 2(b). $C_4^-$ has consistently higher probability of emission than $C_1^-$ from irradiated SWCNTs in 2(a) while the opposite is true in the case of irradiated MWCNTs in 2(b).

The normalized emission probabilities $p(C_x)$ evaluated for each species from the mass spectra of Fig. 2(a) and (b) from SWCNTs and MWCNTs are shown in Fig. 3(a) and (b) where $p(C_x)$ for $C_1^-$, $C_2^-$, $C_3^-$ and $C_4^-$ are plotted as a function of $E(Cs^+)$. Di- and tri-atomic cluster emissions are the higher probability events in both types of nanotubes. The emission of the $C_4^-$ cluster occurs whenever a quarto-valent defect is formed. Its emission is seen to be more probable than that of $C_1^-$ in the case of SWCNTs as opposed to the emissions from MWCNTs under similar irradiations. The formation of multi-vacant defects with the consequent emissions of clusters $C_x$ has been described due to the initiation of localized thermal spikes in SWCNTs[20]. The localized thermal spikes have been indicated to operate even in the bulk solids of the irradiated Si and Ge single crystals and the molecular solid ZnO[32]. In MWCNTs, due to the multi-shelled layers of the irradiated nanotube, the collision cascades are more pronounced with the emission of monatomic carbon anion $C_1^-$ with higher probability than $C_4^-$ emitted from spike-sublimed surfaces as is evident in Fig. 3(b). Apart from this difference in the emission probabilities of $C_4^-$ and $C_1^-$, cluster anions $C_2^-$ and $C_3^-$ remain the highest probability outputs of the dissipative structures with thermal origin, operating in and on the surfaces of MWCNTs, just as the emissions are from SWCNTs.

### C. Multi-scaled fractal dimension $d_f^x$ and relative entropy $D(p_x \parallel p_y)$

The uncertainty associated with any event or series of events with probability mass function $p(\zeta)$ is defined and measured by the product $p(\zeta)\ln(1/p(\zeta))$. The sum over all $\zeta$ of this product is the well-known Shannon entropy or the information[24]

$$I_\zeta = \sum_\zeta p(\zeta)\ln(1/p(\zeta)) \qquad (2)$$

It is dependent upon the combined ion energy measure $\zeta$ defined above in eq.(1). It depends on $E(Cs^+)$ and $\delta E(Cs^+)$. Fractal or the information dimensional analysis is based on $I_\zeta$. For carbon atoms and clusters emitted from the irradiated nanotubes, $I_\zeta$ is shown to be an indicator of different phases of the emerging dissipative structures. Fractal dimension[33-35] is defined following Renyi[36] as

$$d_f^x = \sum_\zeta p_x(\zeta) \ln(1/p_x(\zeta))/\ln(1/\zeta) = I_\zeta/\ln(1/\zeta) \qquad (3)$$

Here $p(\zeta)$ is the respective probability distribution of the *x*-th component of the emitted species $C_x$ and $\zeta$ is the number of distributive stages or the measure required to obtain information $I_\zeta$.

Another information-theoretic parameter, relative entropy is introduced and calculated for the dynamical systems to provide a measure of the Kullback-Leibler distance[25] between two probability distributions $p_x \equiv p_x(\zeta)$ and $p_y \equiv p_y(\zeta)$ that represent the probability distributions of any two sputtered species $C_x$ and $C_y$. Relative probability is evaluated as[25-28]

$$D(p_x \parallel p_y) = \sum p_x(\zeta) ln(p_x(\zeta)/p_y(\zeta)) \quad (4)$$

Relative probability is generally asymmetric between the two random probability distributions implying $D(p_x \parallel p_y) \neq D(p_y \parallel p_x)$. The experimentally determined probability distributions for the sputtered atoms and clusters species have been used to calculate the two information-theoretic functions $d_f^x$ and $D(p_x \parallel p_y)$ and shown in Table 1 for the coarse-grain $\delta E(Cs^+) = 0.5\ keV$ and the finer-grain $\delta E(Cs^+) = 0.1\ keV$.

In the column of $d_f(C_1)$ in Table 1(A), the values for SWCNTs and MWCNTs are $\lesssim 1$ for all of the chosen energy ranges of $E(Cs^+)$ and for the two measures with varying scales $\delta E(Cs^+)$. Fractal dimension of monatomic carbon $d_f(C_1)$ is 0.44 for SWCNTs and 0.99 for the MWCNTs. The fractal dimension of C$_2$ $d_f(C_2)$ is between 1.29 and 1.82 for the three different irradiation experiments. The case for triatomic cluster C$_3$ has average value of $d_f(C_3) = 1.44$ for the seven combinations of $E(Cs^+)$ and the two nanotubes.

**Table I. $d_f^x$ and $D(p_x \parallel p_y)$ for the coarse-grain $\delta E(Cs^+) = 0.5\ keV$ and the fine-grain $\delta E(Cs^+) = 0.1\ keV$.**
(A) Fractal dimension $d_f(C_x) \equiv d_f^x$ for the sputtered species $C_x$ are tabulated for SWCNTs, MWCNTs irradiated with Cs$^+$ in the three energy ranges. The fractal dimensions of C$_1$, C$_2$ and C$_3$ sputtered from SWCNTs and MWCNTs under similar irradiation conditions are compared.

| (A) | $d_f(C_1)$ | $d_f(C_2)$ | $d_f(C_3)$ |
|---|---|---|---|
| *(a) $\delta E(Cs^+) = 0.5\ keV$* | | | |
| *E(Cs$^+$)=0.5-3.5 keV* | | | |
| SWCNT (2nm Ø) | 0.91 | 1.29 | 1.32 |
| MWCNT (6nm Ø) | 1.04 | 1.35 | 1.38 |
| | | | |
| *(b) $\delta E(Cs^+) = 0.1\ keV$* | | | |
| *E(Cs$^+$)=0.3-1.0 keV* | | | |
| SWCNT (2nm Ø) | 0.44 | 1.33 | 1.39 |
| MWCNT (6nm Ø) | 0.99 | 1.57 | 1.63 |
| *(c) $\delta E(Cs^+) = 0.1\ keV$* | | | |
| *E(Cs$^+$)=0.8-2.0 keV* | | | |
| SWCNT (pristine) | 1.01 | 1.82 | 1.66 |
| SWCNT (heavily irradiated) | 0.67 | 1.8 | 1.43 |

**(B)** Relative entropies $D\left(p(C_x) \| p(C_y)\right)$ are shown for the two sets of sputtered species ($C_2$, $C_1$) and ($C_2$, $C_3$) as the 4 columns under $D(p(C_2) \| p(C_1))$, $D(p(C_1) \| p(C_2))$, $D(p(C_2) \| p(C_3))$ and $D(p(C_3) \| p(C_2))$.

| (B) | $D(p(C_2) \| p(C_1))$ | $D(p(C_1) \| p(C_2))$ | $D(p(C_2) \| p(C_3))$ | $D(p(C_3) \| p(C_2))$ |
|---|---|---|---|---|
| *(a) $\delta E(Cs^+) = 0.5\ keV$* *$E(Cs^+)$=0.5-3.5 keV* | | | | |
| SWCNT (2nm Ø) | 4.22 | 2.95 | 1.54 | 1.35 |
| MWCNT (6nm Ø) | 2.85 | 2.07 | 0.76 | 0.69 |
| *(b) $\delta E(Cs^+) = 0.1\ keV$* *$E(Cs^+)$=0.3-1.0 keV* | | | | |
| SWCNT (2nm Ø) | 8.83 | 1.79 | 1.97 | 1.79 |
| MWCNT (6nm Ø) | 3.89 | 0.61 | 0.66 | 0.62 |
| *(c) $\delta E(Cs^+) = 0.1\ keV$* *$E(Cs^+)$=0.8-2.0 keV* | | | | |
| SWCNT (pristine) | 8.46 | 5.48 | 2.99 | 2.55 |
| SWCNT (heavily irradiated) | 14.91 | 6.98 | 3.17 | 2.72 |

Table I(B) tabulates the relative entropy $D\left(p(C_x) \| p(C_y)\right)$ between the two probability distributions $p(C_x)$ and $p(C_y)$ that give the numerical estimates of the Kullback-Leibler distance. In the experiments reported here, it establishes the separation between the physical processes that are responsible for any of the two outputs $C_x$ and $C_y$. Rather than the numerical value of $D\left(p(C_x) \| p(C_y)\right)$, it is the relative difference between the sets of the asymmetric relative entropies that defines the nature and extent of the physical separation. Hence, we calculate $D\left(p(C_x) \| p(C_y)\right)$ and $D\left(p(C_y) \| p(C_x)\right)$ to compare and evaluate the actual separation, as a measure of distinguishing the different dissipative structures. The minimum numerical difference between the two relative entropies of $C_2$ and $C_3$ is 3% while the maximum is 14%. The two clusters ($C_2$ and $C_3$) are the byproducts of the same dissipative structure generated by the localized thermal spike[20,28,32]. The monatomic $C_1$ and $C_2$ are the flag-bearers of the two distinct physical mechanisms and therefore, their relative entropies $D(p(C_2) \| p(C_1))$ and $D(p(C_1) \| p(C_2))$ show marked

differences in their respective magnitudes. Their mutual ratio is shown as 4.93 for SWCNTs and 6.4 for the MWCNTs obtained with the smallest $\delta E(Cs^+) = 0.1\ keV$. It reduces to about 1.4 for $\delta E(Cs^+) = 0.5\ keV$.

**D. Spatio-temporal description of the dissipative structures of cascades and spikes**

The experimental observations from the mass spectra of atoms and clusters sputtered from single and multi-walled carbon nanotubes for different ranges of Cs$^+$ energies and the two $\delta E(Cs^+)$ measures show that (a) the dominant emission of $C_2$, $C_3$, $C_4$ and higher clusters is a persistent feature of the mass spectra over the monatomic emissions, (b) the absence of $C_1$ in the mass spectra at very low Cs$^+$ energies and its consistent, low relative intensities as compared with those of the clusters are coupled with an energy dependence of the probability p($C_1$) for the emission of $C_1$ on Cs$^+$ energy $p(C_1) \propto (dE/dx)^\gamma$. Collision cascade theories[29,30] and the Monte Carlo simulations SRIM[10] provide elaborate theoretical and simulation-based justifications for the mechanisms of energy loss $(dE/dx)^\gamma$ in the bulk solid media, the exponent $\gamma$ includes the interatomic potentials and the assumptions about the binding energies etc. SRIM has been a powerful tool for the bulk media and a wide range of ionic types and energies. It however, does not, and we have shown elsewhere, that it cannot, predict the thermal spikes either in the bulk solids or the sp$^2$-bonded single and multi-shelled structures of carbon. The probabilities of emission of $C_2$, $C_3$ and $C_4$ show an energy-independence that depend upon the conditions for the generation of a spike in carbon hexagonal patches[20,28]. The probability of emission of a cluster $C_x$ with energy of formation $E_{xv}$ for an x-valent vacancy formation at temperature $T_{spike}$ is estimated as $p(C_x) \propto \left(exp(E_{xv}/T_{spike}) + 1\right)^{-1}$; $p(C_x)$ is not dependent on Cs$^+$ energy but on $T_{spike}$.

Figure 4 is the schematic representation of the initiation of the energy dissipative processes in the irradiated sp$^2$-bonded atoms of the hexagonal networks of single and multi-walled carbon nanotubes. Different energy and time scales are associated with the two processes. The first shown in Fig. 4(a) is initiated with energies received by the atoms of the hexagons $E_{cascades} \geq E_{disp}$, where $E_{disp}$ is the energy required to displace an atom from its site[10]. In SWCNTs and MWCNTs, $E_{disp} \sim 30 - 40 \, eV$. Binary atomic collision cascades are initiated with energies $\geq E_{disp}$ and occur at time scales $\sim 10^{-15} - 10^{-14} \, s$; the lower limit depends upon the ionic energy and the higher on the primary knock-ons. Localized thermal spikes are initiated with energies $\ll E_{disp}$ where the sharing and recycling of energy in the hexagonal patches as shown in Fig. 4(b). It start around $10^{-13}$ s and subsides into collective atomic vibration time scale $\sim 10^{-12} \, s$. The two processes occur in the same spatial regions but happen at different time scales. Emission of $C_1$ from the collision cascades in Figure 4(a) represents a non-equilibrium, linear process with $d_f(C_1) \sim 1$ that occurs at T~300K. It is the by-product of a high energy ($\geq E_{disp}$), low information-theoretic entropy, dissipative structure with durations $\sim 10^{-14}$ s.

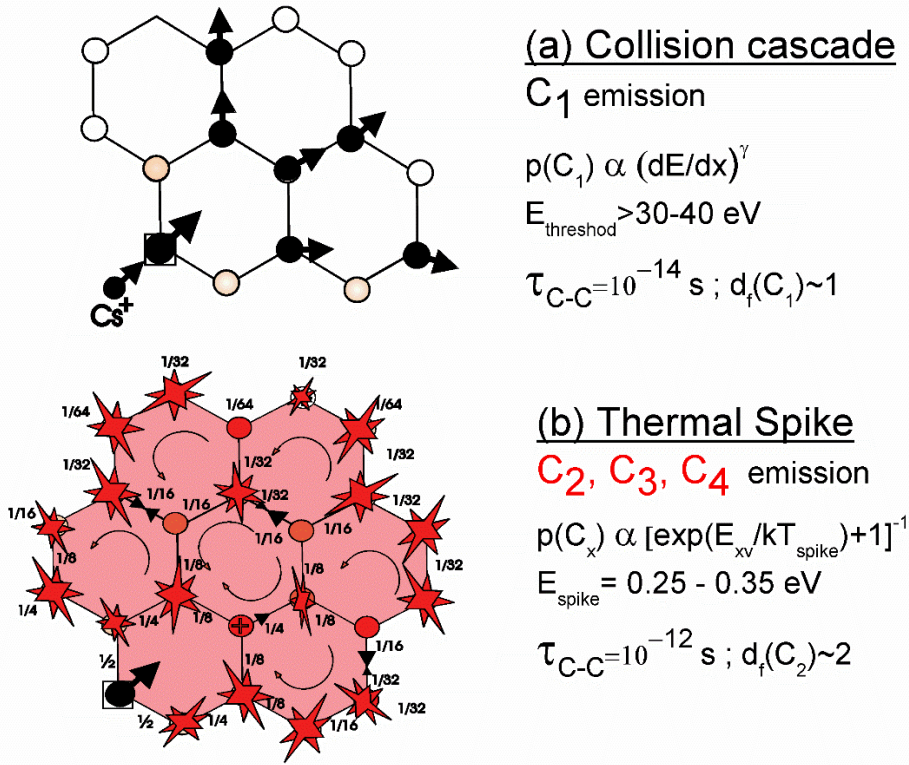

FIG.4. $Cs^+$-initiated dissipative processes in hexagonal patches. (a) For energy received by atom $>E_{disp}$ cascades are likely to start, creating single vacancies, sputtering atoms $C_1$. Probability of emission depends on linear energy dissipation $(dE/dx)^\gamma$ where $\gamma$ identifies the atomic collision mechanisms. The collision

times $\tau_{C-C} \sim 10^{-14}$ s; $d_f(C_1) \sim 1$ (b) Thermal spikes generated by recycling of energy by atoms of the adjacent hexagons for $\sim 1$ eV, leading to localized spikes. Probabilities of cluster emission $p(C_x) \propto \left(exp(E_{xv}/T_{spike}) + 1\right)^{-1}$, depend upon spike temperature $T_{spike}$. Spikes occur for longer times $\tau_{C-C} \sim 10^{-12}$ s; fractal dimension of the representative cluster $C_2$ is $d_f(C_2) \sim 2$.

Clusters $C_2$, $C_3$ and higher ones are emitted from thermal spikes in Figure 4(b) that represent regions in localized thermal equilibrium at $T_{spike} \approx T_{sublimation}$. These are low energy (~0.3 eV) events that occur in the relatively higher entropy-generating dissipative structures with time duration in the inter-atomic collisions $\tau_{C-C} \sim 10^{-12}$ s. The information theoretic entropy and the relative entropy of the probability distributions belonging to the two different physical processes and the data on fractal dimensions of the emitted species has been used here to determine the physically distinguishable features of the cascades and spikes. The graphical representation of the route to the generation of hot, subliming patches that emerge due to the energy recycling hexagons. These lead to the localized thermal spikes[20] where $T_{spike} \sim 3500 - 4000 K$. The fractal dimensions of the emitted species in Table I are shown to indicate the existence of two different energy dissipation mechanisms; one that produces $C_1$ as the output and the other where clusters are emitted. Clusters, in any case, cannot be produced by the linear atomic collision sequences that generate defects in the form of sputtered or interstitial atoms and mono-valent vacancies[8,10,12]. Cluster emission can only be described by the localized thermal spikes[20,28]. Identification and diagnosis of the simultaneous existence of two dissipative processes that are generated by the same incident $Cs^+$ ion is described in this communication by the information-theoretic technique of multi-scaling of fractal dimensions and relative entropies.

## IV. Conclusions

Collision cascades and thermal spike initiated in single and multi-shelled carbon nanotubes are shown to emerge as dissipative structures with the $Cs^+$-induced energy with the energy step $\delta E(Cs^+)$ as the input signal. The energy is consumed and dissipated in linear and nonlinear processes initiated in the nanotubes. The output signal is in the form of sputtered atoms and clusters, emitted from the irradiated single and multi-walled carbon nanotubes. The probability distributions $p(C_x)$ of the sputtered species are constructed for every species $C_x$ emitted from the irradiated nanotubes as function of the multi-scaling measure $\zeta = \zeta(E(Cs^+), \delta E(Cs^+))$. The information-theoretic entropy $I_\zeta(C_x)$ is compiled for each of the emitted constituent. The collision cascades and thermal spikes emerge with distinct temporal and spatial profiles. The spatial profiles of the emitted species are identified by the fractal dimension $d_f(C_x)$ derived from respective $I_\zeta(C_x)$ while the relative entropy $D\left(p(C_x) \parallel p(C_y)\right)$ provides a measure of the distance (or connectivity) between the dissipative structures. The multi-scaling property is shown to identify the roles played by, and the contributions made to, the different dissipative structures operating in the same dynamical system. In the series of experiments reported here the technique of coarse-graining was employed on the irradiated single and multi-walled carbon nanotubes. We have shown that when employed together, the fractal dimension and relative entropy can unambiguously diagnose and characterize the dissipative structures with the multi-scaling approach employed by going from a relatively coarse-grain to a finer-grained scale. The model can be extended to other physical and chemical environments to diagnose and ascertain the nature of the competing dissipative structures.

**ACKNOWLEDGMENTS**

Authors are grateful to the technical staff of 2 MV Pelletron at CASP, Government College University, Lahore during experiments with SRIM ion source.